\documentclass[aps,showpacs,prd,twocolumn,superscriptaddress,floatfix]{revtex4}
\usepackage{amsmath, amsfonts, hyperref}
\usepackage{graphicx}
\newcommand{\beq}{\begin{equation}}
\newcommand{\eeq}{\end{equation}}
\newcommand{\beqn}{\begin{eqnarray}}
\newcommand{\eeqn}{\end{eqnarray}}

\newcommand{\apjl}{Astrophys.\ J.\ Lett.}
\newcommand{\mnras}{Mon.\ Not.\ R.\ Astro.\ Soc.}

\begin{document}
\title{Self-interacting dark matter cusps around massive black holes}
\date{\today}
\author{Stuart~L.~Shapiro}
\altaffiliation{Also Department of Astronomy and NCSA, University of
  Illinois at Urbana-Champaign, Urbana, Illinois 61801, USA}
\author{Vasileios Paschalidis}
\affiliation{Department of Physics, University of Illinois at
  Urbana-Champaign, Urbana, Illinois 61801, USA}

\begin{abstract}
We adopt the conduction fluid approximation to model the steady-state 
distribution of matter around a massive black hole at the center of a 
weakly collisional cluster of particles. By ``weakly collisional" we mean
a cluster in which the mean free time between particle collisions is much longer
than the characteristic particle crossing (dynamical) time scale, 
but shorter than the 
cluster lifetime.  When applied to a star cluster, we reproduce 
the familiar Bahcall-Wolf power-law cusp solution for the stars 
bound to the black hole. Here the star density scales with
radius as $r^{-7/4}$ and the velocity dispersion as $r^{-1/2}$ 
throughout most of the gravitational well of the black hole.
When applied to a relaxed, self-interacting dark matter (SIDM) halo with a 
velocity-dependent cross section $\sigma \sim v^{-a}$, the gas again 
forms a power-law cusp, but now the SIDM density
scales as $r^{-\beta}$, where $\beta = (a+3)/4$, while its velocity
dispersion again varies as $r^{-1/2}$. Results are obtained 
first in Newtonian theory and then in full general relativity. Although
the conduction fluid model is a simplification, it provides a 
reasonable first approximation 
to the matter profiles and is much easier to implement than a full 
Fokker-Planck treatment or an $N$-body simulation of the Boltzmann equation
with collisional perturbations.

\end{abstract}
\maketitle

\section{Introduction}

Determining the stellar distribution around a massive black hole at the
center of a virialized star cluster is an interesting and  well-studied problem.
Originally formulated by Peebles~\cite{Pee72a, Pee72b}, the problem 
was solved for the 
profile in a spherical globular cluster by Bahcall and Wolf~\cite{BahW76}. They 
assumed an isotropic velocity distribution function to solve numerically 
the one-dimensional Fokker-Planck equation for the steady-state 
stellar distribution function $f(E)$ of stars bound to the black hole.
Here $E$ is the energy per unit mass of a star.
They found that the cumulative, distant, two-body encounters
(i.e. small-angle Coulomb scattering) 
between stars drives the density throughout most of the gravitational well
of the black hole (the ``cusp'') to a power-law profile that varies with 
radius as $\rho \propto r^{-7/4}$, while the velocity dispersion
varies as $v \propto r^{-1/2}$.
The effects of velocity anisotropy and the role of the stellar 
disruption loss cone in the cusp were delineated by 
Frank and Rees~\cite{FraR76} and Lightman and Shapiro~\cite{LigS77}. 
They showed that while the distribution function depends
only logarithmically on $J$, the angular momentum per 
unit mass of a star, the inward flux of stars and stellar disruption rate 
by the black hole are affected more significantly. They also applied
the results to cusps in dense galactic nuclei as well as star clusters. 
All of this work motivated detailed two-dimensional
Fokker-Planck treatments to determine $f(E,J)$, both by
Monte Carlo~\cite{ShaM78, MarS79} and by 
finite-difference~\cite{CohK78} methods. 
The role of the removal of bound stars by the black hole 
in heating the ambient star cluster and reversing
secular core collapse (i.e. reversing the ``gravothermal catastrophe'') 
was pointed out by Shapiro~\cite{Sha77} and studied numerically
via time-dependent Monte Carlo simulations of the 
two-dimensional Fokker-Planck equation 
to determine  $f(E,J;t)$~\cite{MarS80}. 
(For a review of some of this early work and references see~\cite{Sha85}).
The same problem has received considerable attention in recent years by many
authors, focusing on such aspects as nonspherical
clusters [e.g.,~\cite{VasM13}, 
mass segregation~\cite{AleH09}, resonant relaxation~\cite{RauT96},
relativistic corrections, tidal disruptions, and extreme-mass 
ratio inspirals (EMRIs) (see~\cite{Ama12} for a review and references)].
  
In this paper we return to the
the steady-state distribution of matter around a massive black hole
at the center of a virialized, weakly collisional, spherical gas, 
such as a star cluster. Here we show that the 
conduction fluid model provides a 
straightforward means of deriving
the steady-state matter density and velocity dispersion profiles in
the cluster.
It is approximate in that it treats the lowest order 
moments of the distribution function,
and not the distribution function itself, and adopts several 
physically reasonable, but simplifying, relations 
to close the moment equations. However, the approach is
much easier to implement than a full Fokker-Planck treatment or an $N$-body 
simulation of the 
Boltzmann equation with collisional perturbations. We then use this approach
to determine the matter distribution in a cusp formed well inside the
homogeneous core of a self-interacting dark matter (SIDM) halo 
containing a massive, central black hole. 

SIDM with cross sections per unit mass in the range  
$\sigma/m \approx 
0.45-450 ~{\mbox{cm}^2\;\mbox{gm}^{-1}}$
[or $8 \times 10^{-(25-22)} ~{\mbox{cm}^2\;\mbox{GeV}^{-1}}$]
was proposed by Spergel and Steinhardt \cite{SpeS00} 
to rectify the problem that cosmological 
simulations~(e.g., \cite{NavFW97, BulKSSKKPD01, WecBPKD02}) 
of purely collisionless cold dark matter (CDM)
exhibit dark matter halos 
that are highly concentrated and have density cusps, 
in contrast to the constant density cores observed for galaxies 
and galaxy clusters. 
 
The SIDM remedy fell out of favor for various reasons, including
possible alternative explanations for this apparent density 
discrepancy (e.g. feedback and expulsion of baryonic gas), 
Bullet cluster observations that seemed to 
constrain SIDM models to have a 
cross section too small to have a significant effect on the
structure of dark matter halos, gravitational lensing and x-ray
data suggesting that the cores of galaxy clusters are denser and less
spherical than predicted for SIDM, as well as 
theoretical biases.
However, the interpretation of some of the data is not definitive 
(see \cite{RocPBKGOM13} for a discussion and references), while   
new observational discrepancies with dissipationless
CDM models have become apparent, such as the absence of dwarf
spheroidal galaxies 
(dSphs) predicted by CDM simulations (the ``missing satellite problem''). 
Moreover, recent SIDM simulations~{\cite{RocPBKGOM13, PetRBK13}
now suggest that SIDM systems
with a (velocity-independent) cross section as large as 
$\sigma/m  \approx 0.1 ~{\mbox{cm}^2\;\mbox{gm}^{-1}}$ can
yield the core sizes and constant central densities observed in 
dark matter halos at all scales, from clusters, to low surface
brightness spirals (LSBs) and dSphs, and are consistent 
with revised Bullet cluster observational constraints. All of the other
large-scale triumphs of CDM appear to be matched by these SIDM simulations.
In addition, simulations involving velocity-dependent SIDM cross sections 
also appear to be successful in forming small, less concentrated cores and 
altering the inner density structure of subhalos in a way
that is compatible with observations of
dSphs but having a negligible effect on galaxy 
cluster scales~{\cite{FirDACH00, FenKTY09, BucF10, LoeW11, VogZL12}.
Such velocity-dependent interactions arise in ``hidden'' sector extensions 
to the Standard Model that have been constructed to explain some 
charged-particle cosmic ray observations in terms of dark 
matter annihilations~{\cite{PosRV08, ArkFSW09, FoxP09, FenKY10a}.
 
The renewed viability of SIDM models motivates our own interest in returning
to a subject that we studied earlier. We previously explored the nature of
the ``gravothermal catastrophe'' (secular core collapse) in isolated SIDM
halos~\cite{BalSI02} using the conduction fluid formalism
(see also~\cite{KodS11} and references therein)
and then showed that the formation of a black hole in the center of a 
galactic halo may be a natural and inevitable consequence of
gravothermal evolution~\cite{BalS02}.

Supermassive black holes (SMBHs)
with masses in the range  $10^6-10^{10} M_{\odot}$ are believed to
be the engines that
power active galactic nuclei and quasars.
There is also substantial evidence that SMBHs reside at the centers
of many, and perhaps most, galaxies~\cite{Ric98, Ho99},
including the Milky Way~\cite{GenEOE97, EckGOS02, GheSHTLMBD05}. 
While SMBHs may arise from baryonic processes, such as the collapse of Pop III 
stars~\cite{MadR01} or supermassive stars~\cite{ShiS02},
followed by mergers and gas accretion, their origin is not yet known.  

It is therefore not unreasonable to imagine
that bound structures containing dark matter of {\it all} sizes  
(galaxy clusters, galaxies, satellite systems, etc) contain massive, central
black holes. This possibility motivates the study in this paper,
for which we adopt a velocity-dependent SIDM cross section $\sigma \propto 
v^{-a}$ and consider a relaxed, spherical, SIDM core with a massive
black hole at its center and determine the cusp profile well inside the core. 
We note that a cusp can also form in 
a purely collisionless CDM system containing
a central black hole. For example,
if a central black hole grows 
adiabatically (e.g. by gas accretion) it 
will perturb the CDM particle orbits while holding the adiabatic invariants
of the motion constant, and this process will result in a 
cusp~\cite{Pee72a, GonS99, SadFW13}. For a SIDM 
system, however, particle collisions will wash out any initial and/or
adiabatically altered particle distribution in a collisional 
relaxation time scale
and the cusp will relax to the solution determined below. 

In Sec. II we summarize our basic model and the assumptions 
that define it. In Sec. III we present our formulation of the
problem in Newtonian theory and use simple scaling to derive
the power-law cusp solution for bound particles obeying a 
power-law velocity-dependent interaction cross section (SIDM matter or stars). 
We also calculate the energy flux conducted
outward by the particles bound to the black hole in the cusp 
and discuss how this flux may
eventually halt and reverse secular core collapse.
In Sec. IV we reformulate the problem in general relativity
and integrate the resulting equations numerically 
to obtain the full density and velocity
profiles in the cusp. In Sec. V we discuss our findings, 
evaluate the contribution of unbound particles, and
consider the very different behavior expected
at very late times when a SIDM core evolves to a collisional fluid.
We also summarize some future work.

We adopt geometrized units and set $G=1=c$ below.

\section{Basic Model and Assumptions}
\label{sec:assump}
Our assumptions apply both to stars in a star cluster and to dark
matter in a SIDM halo; in either case we refer to the matter as
``particles.''  We adopt assumptions similar to those made by Bahcall
and Wolf~\cite{BahW76} who treated stars in a globular star cluster
containing a central black hole:

\begin{enumerate}

\item \label{it:one} 
The distribution of particles is spherically symmetric in space,
isotropic in velocity and has relaxed to a near-equilibrium state. 

\item \label{it:oneA}
Outside the central cusp about the black hole the cluster has a 
nearly homogeneous central core in which the particle rest-mass
density $\rho_0$ and (one-dimensional) velocity dispersion $v_0$ are nearly
constant.

\item \label{it:two} 
The central Schwarzschild black hole has a  
mass $M$ that is much less than the mass 
of the cluster core but dominates the total mass of all particles 
bound to it in the cusp.

\item \label{it:three}
The particles all have the same mass $m$, which is very small compared
to $M$.

\item \label{it:four} 
The mean free path of the particles with respect to self-interactions
is much longer than their characteristic radius from the black hole. 

\item \label{it:five} 
The relaxation time scale due to self-interactions 
between particles is shorter than 
the age of the cluster.

\item \label{it:six} 
A particle is removed from the distribution at a sufficiently small
radius deep inside the gravitational potential well of the black hole,
but outside its horizon.

\end{enumerate}

The characteristic gravitational capture radius of the black hole
$r_{\rm h}$ is taken to be 
\begin{equation}
\label{eq:rh}
r_{\rm h} = \frac{M}{v_0^2},
\end{equation}
The cusp is the region $r \lesssim r_{\rm h}$ inside the core.
Particles whose orbits lie
entirely in the cusp are bound to the black hole.

By assumptions \ref{it:four} and \ref{it:five} 
the particle distribution function satisfies the collisionless 
Boltzmann equation in steady state to a high degree, but the particular 
solution to which it relaxes is determined by the perturbations 
induced by collisions.
By taking suitable velocity moments, the collisionless Boltzmann equation 
reduces to fluid conservation equations for an ideal gas if the velocity
distribution of the particles is isotropic. 
The conduction fluid model provides an approximate means handling this 
system, closing the moments  
to third order when collisions are relevant. This formalism was 
adopted by Lynden-Bell and Eggleton~\cite{LynE80} to  treat the 
gravothermal catastrophe in star clusters (see also~\cite{Spi87}) and 
by Balberg {\it et al.}~\cite{BalSI02} to analyze the gravothermal 
catastrophe in isolated SIDM halos
(see also~\cite{AhnS05}, where the validity of the conduction fluid
approximation is discussed, and ~\cite{KodS11}, where it was shown
to agree with $N$-body simulations that incorporated collisional
perturbations to treat SIDM systems). 

In a star cluster relaxation is driven by multiple, 
small-angle gravitational (Coulomb) encounters. The local relaxation
time scale is given by (see e.g., \cite{LigS78,Spi87})

\begin{eqnarray}
\label{tr_stars}
t_r({\rm stars}) &=&  
%\frac{3^{3/2} v^3}{15.4 G^2 m \rho \ln(0.4 N)},  \cr
\frac{3^{3/2} v^3}{15.4 m \rho \ln(0.4 N)},  \cr
&\simeq& 0.7 \times 10^{9}\mbox{yr} \left(\frac{v}{\mbox{km sec}}\right)^3 \cr
&\times& \left(\frac{M_{\odot}{\mbox{pc}^{-3}}}{\rho}\right)
\left(\frac{M_{\odot}}{m}\right)
\left( \frac{1}{\ln(0.4N)}\right),
\end{eqnarray}
where $v(r)$ is the local one-dimensional 
velocity dispersion, $\rho(r) = mn(r)$ the local (rest) mass density, 
$n(r)$ is the local stellar number density, and
$N$ is the total number of stars in the cluster. The three-dimensional velocity 
dispersion is $v_m(r)=3^{1/2} v(r)$, by isotropy.

In a SIDM halo relaxation is driven by close, large-angle,
elastic interactions between particles. The relaxation time scale 
is the mean time between single collisions and is given by
\begin{eqnarray}
\label{tr_sidm}
t_r({\rm SIDM}) &=& \frac{1}{\eta \rho v \sigma} \cr
&\simeq& 0.8\times 10^{9}\mbox{yr}
             \left[\left(\frac{\eta}{2.26}\right)
             \left(\frac{\rho}{10^{-24}\mbox{g cm}^{-3}}\right)\right. \cr
&\times &\left. \left(\frac{v_0}{10^7\mbox{cm sec}^{-1}}\right)
\left(\frac{v}{v_0}\right)^{1-a}
       \left(\frac{\sigma_0}{1~\mbox{cm}^2\;\mbox{g}^{-1}}\right)\right]^{-1}\,
\end{eqnarray}
where $\sigma = \sigma_0 (v/v_0)^{-a}$ is the cross section per unit mass and
the constant $\eta$ is of order unity. For example, 
$\eta = \sqrt{16/\pi}\approx 2.26$ for particles 
interacting elastically like billiard balls (hard spheres) with a 
Maxwell-Boltzmann velocity distribution
[see \cite{Rei65}, Eqs.~(7.10.3), (12.2.8) and (12.2.12)]. We note that
for a Coulomb-like cross section, where $a=4$,  $t_r({\rm SIDM})$ scales 
the same way with $v$ and $\rho$ as
$t_r({\rm stars})$, up to a slowly varying log factor: 
$t_r \propto v^3/\rho$. We exploit this equivalence below.

A star of radius $R$ 
is tidally disrupted by the black hole
whenever it passes within a radius 
\begin{equation}
\label{eq:rD}
r_D \simeq R (M/m)^{1/3}.
\end{equation}
In a SIDM halo, a particle plunges directly into the black hole once
it passes within a radius 
\begin{equation}
\label{eq:4M}
r_{mb} =4M,
\end{equation}
in Schwarzschild coordinates.
The radius $r_{mb}$ is the radius of the marginally bound circular orbit
with energy (including rest-mass energy) $E/m=1$. 
It is also the minimum periastron of all 
parabolic ($E/m=1$) orbits.
Particles that approach the black hole from large distances 
are typically nonrelativistic 
(i.e. $v_{\infty} \ll c, \ \ E/m \approx 1$) and hence arrive 
on very nearly parabolic orbits. Any such particle
that penetrates within $r_{mb}$ must plunge directly into the black hole
(see, e.g.,~\cite{ShaT83, SadFW13}). 

The net result is that we may set an inner boundary to the cusp at 
$r=r_D$ in a star cluster and at $r=r_{mb}$ in a SIDM halo. At this boundary
the density of particles plummets, so it is a good approximation to 
set the density equal to zero at this radius.
For main sequence stars $r_D \gg M$, so the velocities and
gravitational fields that determine the stellar distribution in the cusp
are entirely Newtonian. By contrast, the
SIDM particle distribution extends into a region in which the spacetime
is highly relativistic.

%Finally we note that as a result of the gravothermal catastrophe
%(secular core collapse) in an SIDM cluster without a black hole,
%the contracting core can evolve to a sufficiently high density that
%the mean free path

\section{Newtonian Model}

In this section we formulate the problem in Newtonian physics, which
will help guide our general relativistic treatment in the next section.
The basic conduction fluid equations required to determine the
secular evolution of bound particles driven by 
collisional relaxation in the cusp are given 
by~\cite{LynE80, Spi87, BalSI02}
\begin{equation}
\label{eq:hydroeq}
%\frac{\partial (\rho v^2)}{\partial r}=-G\frac{M \rho}{r^2}
\frac{\partial (\rho v^2)}{\partial r}=-\frac{M \rho}{r^2}
\end{equation}
\begin{eqnarray}
\label{eq:firstlaw}
\frac{\partial L}{\partial r} &=& - 4\pi r^2 \rho 
\left\{\left.\frac{\partial}{\partial t}\right|_M\frac{3 v^2}{2}+
      p\left.\frac{\partial}{\partial t}\right|_M\frac{1}{\rho}\right\}  \cr 
\cr \cr
&=& 
-4\pi r^2 \rho v^2 \left(\left.\frac{\partial}{\partial t}\right|_M\right)
\ln\left(\frac{v^3}{\rho}\right) \cr
&=& 0. 
\end{eqnarray}

Equation~(\ref{eq:hydroeq}) is the equation of hydrostatic equilibrium,
where the kinetic pressure $P$ satisfies $P=\rho v^2$. 
Equation ~(\ref{eq:firstlaw}) is the energy equation (the first law of
thermodynamics) for the rate of change of the entropy $s$ given
by 
\begin{equation}
\label{eq:entropy}
s=\ln \left(\frac{v^3}{\rho}\right).
\end{equation}
The time derivatives in Eq.~(\ref{eq:firstlaw}) are Lagrangian, but
in steady state the cluster is virialized and at rest on a dynamical 
time scale and the mean fluid velocity is everywhere negligible.
Hence the time derivatives satisfy
$\frac{d}{dt} \approx \frac{\partial}{\partial t}$ and can be set equal
to zero in seeking the steady-state solution (cf. \cite{BahW76, Hydrostat}). 
As a result, the luminosity
$L$ due to heat conduction is constant. Hence Eq.~(\ref{eq:firstlaw})
can be replaced by
\begin{equation}
\label{eq:lcons}
L={\rm constant}.
\end{equation}
By assumption~(\ref{it:four}), $L$ may be evaluated as a conductive
heat flux in the long mean free path limit,
\begin{equation}
\label{eq:flux}
\frac{L}{4 \pi r^2}=-\frac{3}{2}b \rho \frac{H^2}{t_r} 
          \frac{\partial v^2}{\partial r}.
\end{equation}
In writing Eq.~(\ref{eq:flux}) we evaluated
the kinetic temperature of the particles according to 
$k_B T = mv^2$, where $k_B$ is Boltzmann's constant.
The parameter $b$ is constant of order unity 
and $H$ is the local particle
scale height. For a gas of hard spheres with a Maxwell-Boltzmann distribution
the coefficient $b$ can be calculated to good precision from transport theory,
and has the value of $b\approx 25\sqrt{\pi}/32 \approx 1.38$
[see \cite{LifP81}, Chap. 1, Eq.~(7.6), and Problem 3]. 
The scale height for typical bound particles in the cusp
can be estimated as $H \approx r$.
It will turn out that the 
density and velocity profiles will not depend on any of
the constant numerical coefficients associated with the factors appearing
in the equations (e.g., $\eta, b, H/r,$ etc.).
The relaxation time scales $t_r$
are given by Eq.~(\ref{tr_stars}) for stars and
Eq.~(\ref{tr_sidm}) for SIDM. We will use Eq.~(\ref{tr_sidm}) 
for both cases, setting $a=4$ in the self-interaction
cross section $\sigma \propto v^{-a}$ to treat star clusters which relax via
Coulomb encounters. 

Using Eqs.~(\ref{tr_sidm}) and (\ref{eq:flux}),  
Eqs.~(\ref{eq:hydroeq}) and (\ref{eq:lcons}) yield 
two coupled, ordinary differential equations (ODEs) for $v(r)$ and $\rho(r)$.
We introduce the following nondimensional variables:
\begin{equation}
\label{nondim}
\tilde{\rho}=\rho/\rho_0, \ \ \tilde{v} = v/v_0, \ \ \tilde{r} = r/r_h,
\end{equation}
Dropping the tildes $(~\tilde{}~)$ for convenience, 
the nondimensional ODEs become
\begin{eqnarray}
\label{eq:ode1newt}
\frac{dv}{dr} &=& \frac{D}{v^{2-a} \rho^2 r^4}, \\
\label{eq:ode2newt}
\frac{d\rho}{dr} &=& -\frac{\rho}{v^2 r^2} -\frac{2 D}{v^{3-a} \rho r^4},
\end{eqnarray}
where the constants multiplying $L$ have been lumped together 
with $L$ to define a new nondimensional luminosity constant $D \propto L$
[see Eq.~(\ref{eq:fluxnewt1})].  The coupled equations
above must be solved together with two boundary conditions that guarantee that
the  cusp profiles join smoothly onto the ambient core at some radius
$r_0 \gg r_h$. The nondimensional boundary conditions are then 
\begin{equation}
\label{eq:bcnewt}
\mbox{b.c.'s}: \ \ \ \rho = 1 = v, \ \ r = r _0 \gg 1.
\end{equation}
A third boundary condition requires the density to vanish
at the inner edge of the cusp at radius 
$r_{in} = r_D$ or $r_{mb}$, depending
on whether the particles are stars or SIDM. Typically $r_{in} \ll r_h$,
which translates to 
\begin{equation}
\label{eq:evnewt}
\mbox{b.c.}: \ \ \ \rho = 0, \ \ r = r_{in} \ll 1.
\end{equation}
To satisfy condition~(\ref{eq:evnewt})
the constant $D$ must be chosen appropriately, making $D$ an
eigenvalue of the system. 

As a trivial example, consider the case $D=0=L$. Here the solution to 
Eqs.~(\ref{eq:ode1newt}) and (\ref{eq:ode2newt}), imposing
boundary conditions~(\ref{eq:bcnewt}), becomes 
\begin{eqnarray}
\label{eq:iso}
v/v_0 &=& {\rm constant} = 1, \cr
%\rho/\rho_0 &=& {\rm exp}\left( \frac{GM/r - GM/r_0}{v_0^2} \right).   
\rho/\rho_0 &=& {\rm exp}\left( \frac{M/r - M/r_0}{v_0^2} \right).   
\end{eqnarray}
This solution represents an isothermal profile with zero heat flux.
However, it does not satisfy boundary condition~(\ref{eq:evnewt})
and must therefore be rejected. Physically, any plausible solution
should exhibit an increasing velocity dispersion as one moves deeper
into the gravitational well of the black hole, i.e. $dv/dr < 0$. 
According to Eq.~(\ref{eq:ode1newt}) this requires that we search
for solutions with $D < 0$. We must fine-tune the search to find that value
of $D<0$ that enables us to satisfy condition~(\ref{eq:evnewt}) at the desired
radius $r_{in}$. 
  
\subsection{Power-law solution}

We postpone displaying full numerical solutions to the above
system of equations for the cusp profiles until the next section, where
we will formulate and solve the same problem in general relativity.
Here, however, we demonstrate that Eqs.~({\ref{eq:ode1newt}) and
(\ref{eq:ode2newt}) admit power-law solutions that should
apply in the cusp interior, well away from its inner and outer boundaries.

Let $v \propto r^{-\alpha}$ and $\rho \propto r^{-\beta}$ and substitute
into Eqs.~({\ref{eq:ode1newt}) and (\ref{eq:ode2newt}).
Equating powers of $r$ on both sides of these two equations yield 
two equations relating $\alpha$ and $\beta$, which when solved simultaneously
yield $\alpha = 1/2$ and $\beta = (3+a)/4$, or
\begin{equation}
\label{eq:powlaw}
v \sim r^{-1/2}, \ \ \ \rho \sim r^{-(3+a)/4}.
\end{equation} 

We note that for the case $a=4$ applicable to star clusters we recover
the Bahcall-Wolf scaling laws: $v \sim r^{-1/2}, \ \ \ \rho \sim r^{-7/4}$.
Moreover the numerical integration of the coupled ODEs with 
their boundary conditions yields a profile that is in excellent agreement
with the fitting functions provided by Bahcall and Wolf to their
numerical solutions of the Fokker-Planck equation in steady state 
(see Sec.~\ref{sec:GRnum}).

The above power-law solution can also be obtained by applying the
simple scaling argument of Shapiro and Lightman~\cite{ShaL76}, who
argued that in steady state the energy flux of bound particles 
must be constant independent of 
radius throughout the cusp and be transported on a relaxation time scale 
(cf. Eqs.~(\ref{eq:lcons}) and (\ref{eq:flux})):
\begin{equation}
\label{eq:LSscaling}
L \sim \frac{N(r) E(r) }{t_r} = \ {\rm constant},
\end{equation}
where $N \sim \rho r^3  \sim r^{-\beta +3} $ is the number of
particles between $r$ and $2r$,
$E(r) \sim v^2 \sim r^{-1}$ is the characteristic energy of a bound particle 
orbiting in this layer, and 
$t_r \sim 1/(\sigma \rho v) \sim 1/r^{(a-1)/2 - \beta}$.
Inserting the factors in Eq.~(\ref{eq:LSscaling}) and solving for
$\beta$ yields the same result found above, Eq.~(\ref{eq:powlaw}). 
Evaluating the result for a velocity-independent cross section
with $a=0$ gives $\rho \sim r^{-3/4}$, a result found previously~\cite{Pet04}
using the Shapiro-Lightman scaling argument.

\subsection{Energy flux}
\label{sec:energy_flux}

Once the eigenvalue $D$ is determined numerically the 
outward kinetic energy flux conducted by particles
throughout the cusp can be evaluated explicitly.
Restoring units we have
\begin{equation}
\label{eq:fluxnewt1}
L = 12\pi D \eta b \sigma_0 \rho_0^2 M^3 v_0^{-3} = {\rm constant}.
\end{equation}
We may recast Eq.~(\ref{eq:fluxnewt1}) as
\begin{equation}
\label{eq:fluxnewt2}
L = 18 D b \frac{N(r_h) E(r_h)}{t_r(r_h)},
\end{equation}
where
\begin{equation}
\label{eq:fluxnewt3}
\frac{N(r_h) E(r_h)}{t_r(r_h)} = \frac{(4 \pi r_h^3 \rho_0/3 m))
(m v_0^2/2)}{(\eta \rho_0 v_0 \sigma_0)^{-1}}, 
\end{equation}
thereby justifying the scaling argument leading to 
Eq.~(\ref{eq:LSscaling}). 

This kinetic heat flux is transported out into
the core surrounding the cusp by particle scattering.
This energy can have a significant impact on
the evolution of the cluster, eventually halting and ultimately reversing
secular core collapse (the gravothermal catastrophe) in an isolated system, as pointed 
out for star clusters~\cite{Sha77} and confirmed by detailed Monte Carlo 
simulations~\cite{MarS80}. The asymptotic rate of 
reexpansion of the core can be estimated by equating the heating rate
emerging from the cusp into the core as given by Eq.~(\ref{eq:fluxnewt3}) 
to the rate of increase of the core energy $\dot{E_0}$, where 
%$E_0 \sim -G M_0^2/R_0$, 
$E_0 \sim -M_0^2/R_0$, 
$M_0$ is the mass of the core, assumed constant, $R_0$ is the core radius, 
%$v_0 \sim (G M_0/R_0)^{1/2}$ and $\sigma_0 \sim v_0^{-a}$.
$v_0 \sim ( M_0/R_0)^{1/2}$ and $\sigma_0 \sim v_0^{-a}$.
The result is 
\begin{equation}
\label{eq:expand}
%  R_0 \rightarrow t^{\frac{2}{7-a}}
R_0 \rightarrow t^{2/(7-a)}
\end{equation}
Equation~(\ref{eq:expand}) agrees with~\cite{Sha77}  for the reexpansion
of a star cluster ($a=4$) containing a massive central black hole:
$R_0 \rightarrow t^{2/3}$.

\section{General Relativistic Model}
\label{sec:GR}

To treat the cusp we invoke Birkhoff's 
theorem to ignore the exterior core and halo and adopt the
Schwarzschild metric to describe the spherical spacetime in the cusp:
\begin{equation}
\label{eq:metric}
ds^2 = - (1-2M/r) dt^2 + \frac{d r^2}{1-2M/r} + r^2 d\Omega^2.
\end{equation}
In using the above (vacuum) metric we neglect the small contribution to the 
stress-energy tensor of the particles orbiting in the cusp about the black
hole $M$, in accord with assumption (3).

To determine the particle profiles in the cusp requires the relativistic 
generalizations of Eqs.~(\ref{eq:hydroeq}) and (\ref{eq:firstlaw}).
Hydrostatic equilibrium becomes
\begin{equation}
\label{eq:hydroeqGR}
%\frac{dP}{dr} = -(\rho+P)\frac{d \ln |{ \vec{\xi} \cdot \vec{\xi}}|}{dr} = 
\frac{dP}{dr} = -(\rho+P)\frac{d \ln |{ \xi^a \xi_a}|}{dr} = 
 -\frac{\rho+P}{1-2M/r} \frac{M}{r^2},
\end{equation}
%where $\vec{\xi}=\partial/\partial t$ is the time Killing vector, so that
where $\xi^a=\partial/\partial t$ is the time Killing vector, so that
%$|{ \vec{\xi} \cdot \vec{\xi}}| = g_{00}$, and
$|\xi^a \xi_a| = |g_{00}|$, $P$ is the kinetic pressure of the particles and
$\rho$ is their {\it total} mass-energy density. 
The evolution of the entropy per particle 
$s$ is now governed by the first law of thermodynamics together with 
energy conservation along fluid worldlines, 
$u_a \nabla_b T^{ab} =0$, where $T^{ab}$ is the total stress energy of the
system (fluid  plus heat) and $u^a$ is the fluid four-velocity. 
As a result, Eqs.~(\ref{eq:firstlaw})--(\ref{eq:flux}) now become 
\begin{equation}
\label{eq:entropyGR}
d \rho/d \tau - \frac{\rho + P}{n} d n /d \tau = 
nT \frac{ds}{d \tau} = -\nabla_a q^a - a_a q^a = 0,
\end{equation}
where $\tau$ is proper time, 
$n$ is the proper particle number density, $T$ is the particle 
kinetic temperature, $a^a$ is their
four-acceleration, and $q^a$ is the heat flux four-vector (see~\cite{BauS10},
 Eq. 5.103).
We will relate $P$ and $T$ to the particle rest-mass density and velocity
dispersion below. The last equality in Eq.~(\ref{eq:entropyGR})
is imposed by our seeking a steady-state solution.

The only nonzero component of $q^a$ for a virialized gas that is at 
rest in a stationary, spherical gravitational field is $q^r$, which can
be calculated from
\begin{eqnarray}
\label{eq:4flux}
%q_r = -\frac{\kappa}{|g_{00}|^{1/2}} \left(T |g_{00}|^{1/2} \right)_{,r} 
q_r &=& -\frac{\kappa}{|g_{00}|^{1/2}} 
\frac{d \left(T|g_{00}|^{1/2} \right)}{dr} \cr   
    &=& -\kappa \left( \frac{dT}{dr} + 
    \frac{T}{1-2M/r}\frac{M}{r^2} \right)
\end{eqnarray} 
where $\kappa$ is the thermal conductivity [see~\cite{MisTW73}, Eq. (22.16j)]. Similarly,
the only nonzero component of $a^a$ is given by
\begin{equation}
\label{eq:accel}
%a_r = \left(\ln |\xi^a \xi_a|^{1/2} \right)_{,r} =
a_r =\frac{d \ln |\xi^a \xi_a|^{1/2}}{dr}  =
\frac{1}{1-2M/r} \frac{M}{r^2}.
\end{equation}
Using Eq.~(\ref{eq:accel}) in 
Eq.~(\ref{eq:entropyGR}) and integrating gives
\begin{equation}
\label{eq:lconsGR}
r^2 q^r = \frac{C}{(1-2M/r)^{1/2}}, \ \ C = {\rm constant}.
\end{equation}
Equation~(\ref{eq:lconsGR}) is the relativistic analog of
Eq.~(\ref{eq:lcons}).

We determine the effective conductivity $\kappa$ taking the Newtonian limit of 
Eq.~(\ref{eq:4flux}) and equating the resulting expression for 
$q^r$ to $L/4 \pi r^2$ given by Eqs.~(\ref{eq:flux}) and (\ref{tr_sidm}). 
In this limit $q^r \approx -\kappa dT/dr$, $\rho \approx mn$,
and $T \approx mv^2/k_B$, where $k_B$ is 
Boltzmann's constant. Matching yields
\begin{equation}
\label{eq:kappa}
\kappa = A v^{1-a}n^2 r^2, \ \ \ A = 
\frac{3}{2} \eta b m k_B \sigma_0 v_0^a = {\rm constant}.
\end{equation}

Combining Eqs.~(\ref{eq:4flux}) and ({\ref{eq:lconsGR}) gives an
equation for the temperature profile T(r):
\begin{equation}
\label{eq:gradT}
\frac{dT}{dr} = \frac{C}{\kappa r^2 (1-2M/r)^{3/2}} - \frac{T}{1-2M/r}
     \frac{M}{r^2}.
\end{equation}
Equations~(\ref{eq:hydroeqGR}) and ~(\ref{eq:gradT}) provide the two coupled 
equations needed to determine the steady-state
particle profiles in the clusters. To apply them we need to relate
the quantities $P(r)$ and $T(r)$ to the density and velocity dispersion 
in the system. For our simplified treatment based on integrating 
the moments of the Boltzmann equation in the fluid conduction 
approximation it is adequate
to model the particles as a perfect, nearly collisionless, 
relativistic gas where all the particles have the same speed locally  
but move isotropically. At each radius we then may set 
$P \equiv nk_BT = \rho v^2$, 
where $v$ is the one-dimensional velocity dispersion, 
$\rho = \gamma m n$ and
$\gamma = 1/(1-3 v^2)^{1/2}$, which gives $k_B T = \gamma m v^2$. 

To cast the two ODEs in a form that most closely resembles the Newtonian
Eqs.~(\ref{eq:ode1newt}) and (\ref{eq:ode2newt}) 
we define the quantities
\begin{equation}
\label{eq:varGR}
\rho_N \equiv m n, \ \ \ v_N^2 \equiv k_B T/m = \gamma v^2,
\end{equation}
and the nondimensional variables
\begin{equation}
\label{eq:nondimGR}
\tilde{\rho}_N=\rho_N/\rho_0, \ \ \tilde{v}_N = v_N/v_0, \ \ \tilde{r} = r/r_h, \ \ {\rm etc}.
\end{equation}
In terms of these variables, but again dropping tildes  $(~\tilde{}~)$,
the two nondimensional ODEs that determine the profiles become
\begin{eqnarray}
\label{eq:ode1GR}
\frac{dv_N}{dr} &=& \frac{D}{v_N^{2-a} \rho_N^2 r^4} 
\frac{f^{a-1}(v_N)}{(1-\frac{2v_0^2}{r})^{3/2}} 
-\frac{v_0^2}{2 r^2(1-\frac{2v_0^2}{r})^{1/2}} \ \ \ \  \\
\label{eq:ode2GR}
\frac{d\rho_N}{dr} &=& -\frac{\rho_N \gamma}{v_N^2 r^2} 
 -\frac{2 D}{v^{3-a} \rho_N r^4}
\frac{f^{a-1}(v_N)}{(1-\frac{2v_0^2}{r})^{3/2}}.
\end{eqnarray}
Here the function $f(v_N)$ inverts Eq.~(\ref{eq:varGR}) to 
give $v$ in terms of $v_N$,
\begin{equation}
\label{eq:fvN}
v^2 = v_N^2 f^2(v_N), \ \ \ f^2(v_N) = 
\frac{1}{(1+9 v_0^4 v_N^4/4)^{1/2} + 3 v_0^2 v_N^2/2}.
\end{equation}
Appearing in the function $f(v_N)$ above, the velocity $v_N$ is again 
normalized as in Eq.~(\ref{eq:nondimGR}), with $v_0$ in units of the
speed of light.  Equations~(\ref{eq:ode1GR}) and (\ref{eq:ode2GR}) for the cusp 
must be solved subject to the two outer boundary conditions set by 
the ambient cluster core, which in nondimensional form become
[cf. Eq.~(\ref{eq:bcnewt})]
\begin{equation}
\label{eq:bcGR}
\mbox{b.c.'s}: \ \ \ \rho_N = 1 = v_N, \ \ r = r _0 \gg 1.
\end{equation}
In addition, a third boundary condition at the inner edge
of the cusp must be satisfied [cf. Eq.~(\ref{eq:evnewt})],
\begin{equation}
\label{eq:evGR}
\mbox{b.c.}: \ \ \ \rho_N = 0, \ \ r = r_{in} \ll 1,
\end{equation}
where, nondimensionally, 
\begin{eqnarray}
\label{eq:bc3GR}
r_{in} &=& v_0^2 (R/M)(M/m)^{1/3} \ \ \ {\rm (stars)}, \cr
 &=&  4 v_0^2 \ \ \ \ \ \ \ {\rm (SIDM)},
\end{eqnarray}
[cf.~Eqs.~(\ref{eq:rD}), and (\ref{eq:4M})].

Once again, the third boundary condition is imposed by finding the
appropriate eigenvalue $D < 0$. 

\subsection{GR  vs Newtonian equations}

From the above ODEs it is evident that
the nondimensional GR profiles are determined by one free physical
parameter, $v_0$, the core velocity
dispersion. The value of $v_0$ sets the degree to which the ambient 
core is relativistic, as well as the dynamic range between the outer
boundary of the cusp, $r_h$ and its inner boundary, $r_{in}$. 
In the Newtonian limit, Eqs.~(\ref{eq:ode1GR}) and (\ref{eq:ode2GR}) 
reduce to Eqs.~(\ref{eq:ode1newt}) and (\ref{eq:ode2newt})
and are independent of the value of $v_0$, although the latter parameter 
enters the inner boundary condition through $r_{in}$. 

When $v_0 \ll 1$ the Newtonian equations suffice to determine the
cusp solution for a cluster of normal stars, as the stars never 
move at relativistic velocities or in a strong-field region before
being disrupted. For a SIDM cusp, by contrast, while the bulk of an
SIDM cusp resides in the Newtonian
regime when $v_0 \ll 1$, the innermost particles have orbits which enter 
the high-velocity, strong-field
region outside the central black hole before being captured. 
The GR equations are then necessary to obtain an accurate solution.

When $v_0 \lesssim 1/3$ the core is relativistic and the GR equations
must be used to treat the cusp everywhere, both for a 
star or a SIDM cluster. 
We point out that as $v_0 \rightarrow 1/3$ in the core 
(i.e. three-dimensional
velocity dispersion $v_m \rightarrow 1$) a (nearly) collisionless core-halo 
cluster inevitably becomes dynamically unstable and undergoes catastrophic
collapse on a {\it dynamical} (free-fall) time scale, forming a massive,
central black hole within the ambient halo 
(see, e.g., \cite{ShaT92, BauS10} for
reviews of and references to both analytic theory and simulations). Subsequent
gravitational encounters 
(stars) or collisions (SIDM) will establish the cusp described here
on a relaxation time scale $t_r$ following the collapse.

\subsection{Energy flux}
\label{sec:eflux}

The kinetic heat flux can be calculated from
\begin{equation}
\label{eq:fluxGR1}
\frac{L}{4 \pi r^2} = |q^a q_a|^{1/2}.
\end{equation}
The relativistic analog of Eq.~(\ref{eq:fluxnewt1}) now becomes
\begin{equation}
\label{eq:fluxGR2}
L= \frac{12\pi D \eta b \sigma_0 \rho_0^2 M^3 v_0^{-3}}{1-2M/r}
= \frac{\rm constant}{1-2M/r}.
\end{equation}
Hence, as the radius from the black hole increases, the outward energy 
flux flowing through the cusp and measured by a local, static observer 
decreases faster than $r^{-2}$. The factor $(1-2M/r)$ now appearing on the
right-hand side of Eq.~(\ref{eq:fluxGR2}) accounts
for the gravitational redshift (both in energy and time) as the heat flux
propagates outwards. 

\subsection{Numerical results}
\label{sec:GRnum}

We integrate Eqs.~(\ref{eq:ode1GR}) and (\ref{eq:ode2GR}),
subject to boundary conditions Eqs.~(\ref{eq:bcGR}) and (\ref{eq:evGR}),
for several different choices
of the velocity power laws characterizing the particle interaction
cross section, $\sigma \propto v^{-a}$. We take the nondimensional
outer boundary radius to be $r_0 = 10 r_h$, which
puts it well outside the cusp and into the homogeneous core. 
We assign the inner boundary radius to be $r_{in} = 10^{-3} r_h$
for illustrative purposes. This radius is small enough to give the 
cusp sufficient dynamic
range to confirm the power-law radial dependence anticipated by 
Eq.~(\ref{eq:powlaw}) and to establish scaling laws applicable to 
other choices of $r_{in}$. For typical star clusters and
SIDM systems, the typical values of $r_{in}$ are much smaller. 
Inverting Eq.~(\ref{eq:bc3GR}), 
the core velocity $v_0$ is related to $r_{in}$ by 
\begin{equation}
v_0 = 140 \left(\frac{m}{M_{\odot}}\right)^{1/6} 
        \left(\frac{R_{\odot}}{R}\right)^{1/2}M_3^{1/3}r_{in,3}^{1/2} \ {\rm km/s}   
\end{equation}
for star clusters, and 
\begin{equation}
v_0 = 4.7 \times 10^3 r_{in,3}^{1/2} \ {\rm km/s} 
\end{equation}
for SIDM halos. Here we define $M_3 \equiv M/10^3 M_{\odot}$ and
$r_{in,3} \equiv 10^3(r_{in}/r_h)$.

The results of the integrations are plotted in Fig.~\ref{fig:dens} for
the rest-mass density profile $\rho_N$ in the cusp.
The numerical profiles confirm that the
power-law density profiles given by Eq.~(\ref{eq:powlaw}) 
apply to the bulk of the cusp, i.e. the region well inside the inner 
and outer boundaries.
The profiles can be fit reasonably well to the general analytic expression
\begin{eqnarray}
\label{eq:analdens}
\rho_N/\rho_0 &=& 1 + \xi (r_h/r)^{(3+a)/4}, \ \ \ r \gtrsim r_{in}, \\
       &=& 0, \ \ \ r \lesssim r_{in},\nonumber
\end{eqnarray}
where $\xi$ is of order unity. We plot four cases in  Fig.~\ref{fig:dens}
corresponding to velocity-dependent interaction cross sections with
power-law parameter $a = 0,1,2,$ and $4$. For these four cases
shown in Fig.~\ref{fig:dens}
we plot four analytic curves for $r \gtrsim r_{in}$, setting 
$(a,\xi) = (0,1.5), (1,1.2), (2,1)$, and $(4,1)$ in
Eq.~(\ref{eq:analdens}). These analytic curves are
seen to match the numerical solutions reasonably well.

\begin{figure}
\includegraphics[trim =0.cm 5.0cm 0.cm 3.cm,clip=True,width=9cm]{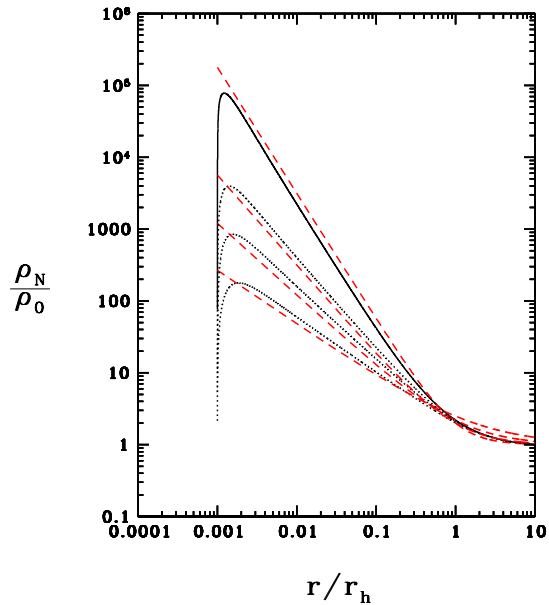}
\caption{
Profiles of the cusp rest-mass density $\rho_N$, normalized to the
core density $\rho_0$, for select choices
of the power-law $a$ in the interaction cross section 
$\sigma \propto v^{-a}$.
Starting from the bottom and moving up, the solid curves show
numerical results for $a = 0, 1, 2$, and $4$. The dashed curves exhibit
crude analytic fits to the numerical profiles that apply 
to all radii $r \gtrsim r_{in}= 10^{-3} r_h$ [Eq.~(\ref{eq:analdens})],  
where $r_h$ is the cusp radius [Eq.~(\ref{eq:rh})].
}
\label{fig:dens}
\end{figure}

The corresponding velocity dispersion profiles for the four cases
are plotted in Fig.~\ref{fig:vel}. They also agree with the power-law
($\sim$ Keplerian) profile predicted by Eq.~(\ref{eq:powlaw}). 
For comparison, the analytic fit
\begin{equation}
\label{eq:analvel}
v/v_0 = 1 + \frac{4}{11}(r_h/r)^{1/2}, \ \ \ r \gtrsim r_{in},
\end{equation}
is also shown on the plot.

%note: ``trim'' works from left, bottom, right, top of figure.
\begin{figure}
\includegraphics[trim =0.cm 5.0cm 0.cm 3.cm,clip=True,width=9cm]{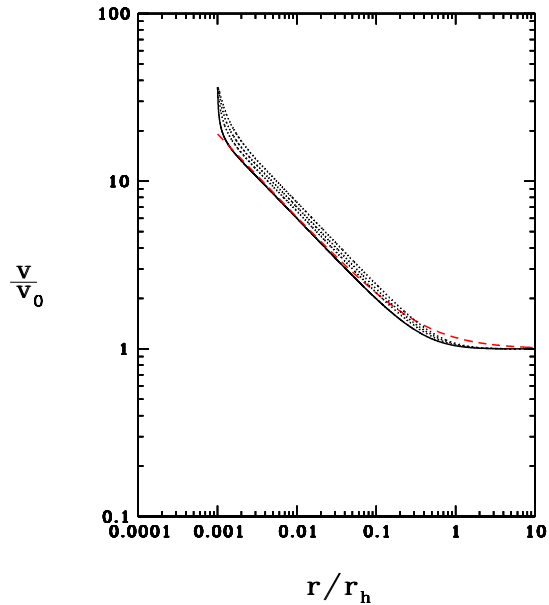}
\caption{
Profiles of the cusp velocity dispersion $v$, normalized to the core
velocity dispersion $v_0$, for select choices
of the power-law parameter $a$ in the interaction cross section 
$\sigma \propto v^{-a}$.
Starting from the top and moving down, the solid curves show
numerical results for $a = 0, 1, 2,$ and $4$. The dashed curve is
a crude analytic fit to the numerical profiles that applies to 
all radii $r \gtrsim r_{in} = 10^{-3} r_h$ [Eq.~(\ref{eq:analvel})],
where $r_h$ is the cusp radius [Eq.~(\ref{eq:rh})].
}
\label{fig:vel}
\end{figure}

The solution of Eqs.~(\ref{eq:ode1GR}) and (\ref{eq:ode2GR}) 
is very sensitive to the eigenvalue $D$. For each choice
of $a$ and $r_{in}$, high integration 
accuracy and iteration are necessary to find the value of $D$ 
to generate a solution that matches the 
required inner boundary condition~(\ref{eq:evGR}). The four cases
examined above yield $(a,D) = (0,~0.7118010817), (1,~0.58832494372), 
(2,~0.50056792714)$ and $(4,~0.3847292732806)$. The 
eigenvalue $D$ determines the outgoing heat flux through
Eq.~(\ref{eq:fluxGR2}).

\section{Discussion}

The case $a=4$ $(\sigma \propto v^{-4})$ 
is particularly interesting, both for star clusters
and SIDM halos. This value corresponds to a Coulomb scattering
cross section and therefore yields the familiar cusp profiles for a 
star cluster containing a massive, central black hole: 
$\rho_N \propto r^{-7/4}$ and $v \propto r^{-1/2}$. Our numerical
results are in close agreement with the steady-state profile found by 
Bahcall and Wolf~\cite{BahW76}, who integrated the Fokker-Planck equation for
a star cluster satisfying the same assumptions adopted here 
(see Sec.~\ref{sec:assump}).  
The analytic fitting formulas defined in Eqs.~(\ref{eq:analdens})
and (\ref{eq:analvel}),
with the very coefficients for these formulas quoted above for $a=4$,
were in fact constructed by 
Bahcall and Wolf to fit their own numerical solution for star clusters.
It is reassuring to 
discover that these formulas also provide good fits to our
numerical solution for
this case, based as it is on a much simpler formulation and calculation. 

Interestingly, the same $a=4$ case is also relevant for some new 
models proposed for SIDM systems. A velocity-dependent, non-power-law, 
elastic scattering cross section, motivated by a Yukawa-like 
potential involving a new  gauge boson of mass 
$m_{\phi}$, (and similar to ``screened''
Coulomb scattering in a plasma) has been
proposed recently~\cite{FenKY10b, FinGSVW11}.
$N$-body simulations adopting this interaction apparently 
can explain the observed cores in dwarf galaxies without
affecting the dynamics of larger systems with larger velocity dispersions,
such as clusters of galaxies~\cite{LoeW11, VogZL12}.

Now the presence of a massive, central black hole in the core of a SIDM system 
will develop a cusp, as we have discussed. For any interaction cross-section,
power law or not, the density and velocity profiles in such a cusp would be 
straightforward to calculate by the method formulated in this paper. 
However, we can already anticipate 
the results for systems interacting via the above Yukawa-like potential, 
provided their core velocity dispersions $v$ satisfy
$\beta_c \equiv \pi (v_{\rm max}/v)^2 \lesssim 0.1$. In this case 
the cross sections vary with velocity 
according to $\sigma \sim \beta_c^2 \sim v^{-4}$,  
up to a slowly varying logarithmic factor.
Here $v_{\rm max}$ is the velocity at which
the momentum-transfer weighted scattering rate $\langle\sigma v\rangle$ peaks at a
cross section value of 
$\sigma^{\rm max} = 22.7/m_{\phi}^2$. When $v_{\rm max}$ is  
comparable to the velocity dispersion in the core of a typical dwarf 
spheroidal galaxy, the $N$-body
simulations that result are reported to give profiles in better agreement 
with observations for these objects than collisionless CDM predictions. 
Adopting such a value for $v_{\rm max}$ implies that throughout 
the bulk of a cusp in dwarf galaxies, as well as in any larger system,
the interaction cross section
resides in the $n=4$ power-law regime, for which the Bahcall-Wolf solution 
applies in a first approximation.

\subsection{Unbound particles}

In addition to the bound particles that orbit in the cusp 
there are {\it unbound} particles 
($E/m > 1$, where $E$ again includes rest-mass energy) 
from the ambient core that penetrate into the cusp. Treating the core 
as an infinite, collisionless 
and monoenergetic bath of such particles with 
rest-mass density $\rho_0$ moving isotropically with (one-dimensional) 
velocity dispersion $v_0$ 
yields a density profile in the gravitational well of the black hole 
given by
\begin{equation}
\label{eq:unbound}
\rho_N/\rho_0 = (1 + 2r_h/3r)^{1/2},
\end{equation} 
and a velocity profile 
\begin{equation}
v/v_0 = (1 + 2r_h/3r)^{1/2},
\end{equation}
assuming Newtonian gravitation~\cite{ZelN71, ShaT83} (for
the solution in general relativity, see \cite{ShaT85}).
The key point is that a comparison of Eqs.~(\ref{eq:powlaw}) 
or (\ref{eq:analdens}) with (\ref{eq:unbound}) shows that
the density of unbound particles penetrating the interior of the cusp is much
smaller than the density that builds up in bound particles for all
plausible ($a>0$) velocity-dependent interactions.

\subsection{Collisional fluid regime}

As shown in~\cite{BalSI02}, gravothermal evolution in a SIDM 
cluster characterized by a velocity-{\it independent} self-interaction 
inevitably drives the contracting core to sufficiently high
density that the particle mean free path to collisions becomes smaller than the
local scale height in the innermost regions. A similar situation 
presumably arises for a range of velocity-dependent interactions. 
When this situation occurs, the particles
in the innermost regions will behave as a collisional (hydrodynamical) fluid. 
A massive black hole at the center of such a core 
would then be expected to accrete particles via steady-state,
adiabatic, spherical Bondi~\cite{Bon52} flow. The cusp in such a case will fill
with gas that has a density profile of the form
\begin{equation}
\label{eq:densbondi}
\rho_N/\rho_0 \approx 1 + \chi (r_h/r)^{3/2},  \ \ \ \chi \sim \mathcal{O}(1), 
\end{equation}
and an inward radial velocity approaching the free-fall
velocity $v_{ff} \sim (M/r)^{1/2}$ inside $r_h$. The 
temperature, or velocity dispersion of the particles, 
$ \gamma v^2 \sim k_B T/m$, will satisfy 
\begin{equation}
\label{eq:tempbondi}
T/T_0 \approx (\rho_N/\rho_0)^{\Gamma-1}. 
\end{equation}
Here the adiabatic index $\Gamma$ is 5/3 for a nonrelativistic 
gas ($k_B T/m \ll 1$) and $4/3$ for a relativistic gas  
($k_B T/m \gg 1$) and varies with radius from
the black hole (see~\cite{ShaT83} for a review of 
spherical Bondi flow, including a
relativistic treatment). For a nonrelativistic
core temperature $T_0$, $\Gamma$ will be near $5/3$ throughout most
of the cusp, decreasing below $5/3$  as the fluid approaches the 
horizon. In this case Eqs.~(\ref{eq:densbondi}) 
and (\ref{eq:tempbondi}) show that the temperature of the fluid 
at the horizon will be marginally
relativistic, climbing to  $k_BT_{\rm horz}/m \sim 1$, independent of $T_0$.

The ratio of the collision mean free path to the characteristic scale height 
in the core of a typical SIDM model constructed to explain currently 
observed clusters satisfies
\begin{eqnarray}
\label{eq:mfp}
\frac{\lambda_0}{R_0} &=& \frac{1}{\rho_0 \sigma_0 R_0} \cr 
  &\approx& 50 \left(\frac{0.1 ~M_{\odot}/{\rm pc}^3}{\rho_0}\right) 
   \left(\frac{1 ~{\rm cm}^2/{\rm g}}{\sigma_0}\right)
   \left(\frac{1 ~{\rm kpc}}{R_0}\right),
\end{eqnarray}
and this ratio increases for $r \gtrsim R_0$. The above inequality is 
consistent with, e.g., the velocity-dependent SIDM models constructed 
in \cite{VogZL12} (see their Fig. 4 for main haloes and Fig. 7 for subhaloes).
Inside a cusp around a central black hole the ratio becomes
\begin{equation}
\frac{\lambda}{r} = \frac{1}{\rho \sigma r}
   \approx \frac{\lambda_0}{R_0} \left(\frac{M_0}{M}\right)
   \left(\frac{r_h}{r}\right)^{(1+a)/4}, \ \ \ r \lesssim r_h,
\end{equation}  
where we have used Eq.~(\ref{eq:powlaw}), and thus 
increases as $r$ decreases below $r_h$ for all $a > -1$. Thus 
a typical SIDM cluster with a central black hole  
is characterized by $\lambda/r \gg 1$ everywhere and
the matter resides in the weakly, and not the fluid, collisional regime. 

Now SIDM clusters {\it can} evolve from a
weakly collisional gas to a strongly collisional fluid, transforming the 
nature of the cusp profiles accordingly. However,
SIDM models constructed to explain observed
clusters are specifically designed {\it not} to have undergone 
such evolution as yet. In particular, these
models employ interaction cross sections whose magnitudes are sufficiently small
that secular core collapse -- the gravothermal catastrophe -- will not
have occurred in a Hubble time. In these cases there is insufficient time 
for the cores to have evolved to a fluid state, which
requires a significant fraction of a core collapse time scale 
($\approx 290 t_r$ for isolated clusters
with a=0; \cite{BalSI02}). This conclusion 
is evident from the relation 
$t_r/t_{\rm dyn} \approx \lambda_0/R_0$, where 
$t_{\rm dyn} \approx 1/\rho_0^{1/2}$ is the characteristic dynamical time scale
in the core. We thus have
$t_r \approx 5\times10^7 (\lambda_0/R_0)(0.1 ~M_{\odot} {\rm pc}^{-3}/
\rho_0)^{1/2}$ yr, whereby substituting Eq.~\ref{eq:mfp}, yields
a typical core collapse time scale significantly longer than the Hubble time.  
Moreover, core collapse is suppressed in (nonisolated) clusters 
when merging from hierarchical formation is included~\cite{AhnS05}.

\subsection{Future work}

The possibility of a massive black hole at the centers of SIDM
clusters and the formation of a cusp around the black hole motivates
several interesting questions that we are now investigating.  One
question is: assuming the SIDM particles experience weak (WIMP-like)
inelastic interactions in addition to the elastic interactions assumed
here, what are the observable signatures of such a cusp?  Will, for
example, the perturbative inelastic interactions produce detectable
radiation or high-energy particles above the values expected from a
homogeneous SIDM core without a cusp~\cite{FiePS13}?  Another question
is, given that the cusp serves as a source of kinetic energy conducted
into the ambient core, is the energy sufficient to reverse secular core
collapse and cause the cluster to reexpand, as discussed in
Sec.~\ref{sec:energy_flux}~\cite{BalS13}?  These are just some of the
issues that we will pursue in future studies of SIDM clusters.

{\it Acknowledgments}: It is a pleasure to thank S. Balberg,
B. Fields, A. Peter and P. Shapiro for useful discussions.  This paper
was supported in part by NSF Grants No. PHY-0963136 and
No. PHY-1300903 and NASA Grants No. NNX11AE11G and No. NN13AH44G at
the University of Illinois at Urbana-Champaign. V. P. gratefully
acknowledges support from a Fortner Fellowship at UIUC.

\bibliography{paper}

\begin{thebibliography}{10}
\bibitem{Hydrostat}%
  \BibitemOpen
  \bibinfo {note} {{The same approach is adopted in constructing a a
  hydrostatic equilibrium model of a slowly evolving star. }}%
  \bibAnnoteFile{NoStop}{Hydrostat}%
\end{thebibliography}
\end{document}